\documentclass[showpacs,preprint,preprintnumbers,amsmath,amssymb,prb]{revtex4}

\usepackage{bm}
\usepackage{epsf}
\usepackage{graphicx}
\usepackage[dvips]{epsfig}

\begin{document}
\title{Carbon nanotube as a nanoscale Cherenkov--type light emitter -- nanoFEL}

\author{K. G. Batrakov, S.A. Maksimenko and P.P. Kuzhir}
\affiliation{\\ Institute for Nuclear Problems, Belarus State
University, Bobruiskaya 11, 220050 Minsk, Belarus}
\author{C. Thomsen}
\affiliation{\\ Institut f\"{u}r Festk\"{o}rperphusik, Technische
Universit\"{a}t Berlin, Hardenbergstr. 36, D-10623 Berlin, Germany}

\begin{abstract}
A  mechanism of stimulated emission of electromagnetic radiation by
an electron beam in carbon nanotubes is theoretically considered.
Three basic properties of carbon nanotubes, a strong slowing down of
surface electromagnetic waves, ballisticity of the electron motion
over typical nanotube length, and extremely high electron current
density reachable in nanotubes, allow proposing them as candidates
for the development of nano-scale Chernekov-type emitters, analogous
to traveling wave tube and free electron laser. Dispersion equations
of the electron beam instability and the threshold conditions of the
stimulated emission have been derived and analyzed, demonstrating
realizability of the  nanotube-based nanoFEL  at realistic
parameters of nanotubes and electronic beams.

\end{abstract}
 \pacs{41.60.-m,78.67.Ch,73.63.Fg}
 \maketitle

\section{Introduction}

Since the discovery of carbon nanotubes (CNTs) in 1991
\cite{Iijima_91}, there has been great interest to their outstanding
structural, electrical and mechanical properties
\cite{Dresselhause_b01,Reich_b04} due to wide applications ranging
from chemical and biological sensors  and actuators to field
emitters to mechanical fillers for composite materials. Among
others, the study of CNTs as building blocks for nanoelectronics
\cite{Collins_00} and nanooptics \cite{Novotny_b06} has continued to
grow unabated owing to the great potentiality for the
miniaturization and the increase of operational speed of
optoelectronic nanocircuits, and for the use in near-field
subwavelength optical element. In that relation, the question of
electromagnetic response properties of CNTs arises. Many interesting
physical effects have been revealed, such as excitation of surface
plasmons \cite{Longe_93}, guiding of strongly slowed-down
electromagnetic surface waves \cite{Slepyan_99_PRB,Maksimenko_b04},
antenna effect -- controlled and enhanced radiation efficiency in
infrared and terahertz ranges
\cite{Hanson_05,Slepyan_06_PRB,Burke_06,Kempa_07,Shuba_07_PRB},
enhanced spontaneous decay rate of an excited atom in the vicinity
of CNT \cite{Bondarev_02_PRL}, and formation of the discrete
spectrum in thermal radiation of finite-length metallic CNTs in the
terahertz range \cite{Nemilentsau_07_PRL}.  Recently,  nanoscale
optical imaging of single-walled CNTs has been studied by means of
high-resolution near-field Raman microscopy
\cite{Hartschuh_PRL_03,Hartschug_JLum_06} and antenna operation of a
CNT array has been demonstrated experimentally \cite{Wang_04_APL}.
Reference  \onlinecite{Rybczynski_07_APL} reports multi-wall CNT as
subwavelength coaxial waveguide for visible light.

An intriguing problem of nanoelectromagnetism is the development of
CNT-based nanoscale sources of light.  A mechanism of the emission
of hard X-radiation by a charged particle moving in a CNT has been
considered  in Ref. \onlinecite{Klimov_96}. The use of CNTs in X-ray
and high-energy particles optics as focusing and guiding elements
and as x-rays sources is presently discussed
\cite{Artru_05_PR,Bellucci_05_NIMB}. In the optical range, the
mechanism of light emission due to exciton recombination in
semiconductor CNTs has been proposed and experimentally verified
\cite{Misewich_03_Sci,Chen_05_Sci}. A possibility of terahertz
emission in CNTs imposed to transverse and axial electric field due
electric-field induced heating of electron gas has been
investigated\cite{Kibis1,Kibis2,Kibis3,Kibis_07_NL}. Recently, the
idea using kinetic energy of CNT-guided electron beam for stimulated
emission of electromagnetic waves in optical and terahertz ranges
has been proposed
\cite{Batrakov_06_SPIE,Kuzhir_07_SRIMN,Batrakov_08_PhE}. In the
given paper we present a consistent theory of the effect.

There is a wide family of devices utilizing interaction of electron
beams with electromagnetic waves to produce electromagnetic
radiation. Started by the invention of klystrons \cite{11}, this
family embraces such well-known systems as traveling wave tubes
(TWT) and backward wave oscillators (BWO) \cite{12}, free electron
lasers (FEL) \cite{Madey,14,Marshall,new}, etc. In systems of that
kind, synchronous motion of electrons and electromagnetic wave
modulates the electron beam and coherent radiation  is produced by
electron bunches. The radiation frequency is smoothly tunable due
its dependence on the electron beam energy. Therefore, such type of
emitters can operate in wide spectral range from microwave and
infrared  frequencies  to vacuum ultraviolet nowadays (e.g., VUV-FEL
at DESY). Several projects aimed with the lasing in hard X-ray range
have started\cite{SLAC,Altarelli}.

The synchronization of moving electrons and electromagnetic wave is
attained either by slowing down the electromagnetic wave (Cherenkov,
Smith-Purcell \cite{19} and quasi-Cherenkov \cite{bat} radiation
mechanisms) or by applying an external magnetic field, which is
uniform in gyrotrons \cite{21,22} and spatially periodical in
undulators \cite{Marshall}. Besides, the oscillator-type mechanism
\cite{fer} is realized for electrons with discrete spectrum of
transverse motion (for example, for  electron channeling in
crystals). The Cherenkov radiation is governed by the
synchronization condition $\omega-\mathbf{k u} =0$, where
$\mathbf{k}$ is the wavevector and $\mathbf{u}$ is the charged
particle (electron) velocity. In systems with external fields the
synchronism condition is transformed to $\omega-\mathbf{k u}-\Omega
=0$ with $\Omega$ as the electron oscillation frequency. In the
oscillator regime $\Omega$ is the transition frequency between
electron levels \cite{Gover}.

For the coherent generation in the devices described, a high vacuum
must be maintained in the region of the electron beam --
electromagnetic wave interaction\cite{miller}. Otherwise, collisions
of electrons with atoms move electrons out the synchronism and,
consequently, lasing is not reached. From this point of view CNTs
are unique objects since they  exhibit ballistic electrical
conduction at room temperature with mean free paths on the order of
microns and even tens of microns
\cite{Frank_98,Berger_02,Berger_03}. Therefore, electrons can emit
coherently from the whole CNT length which is typically 1-10 $\mu$m.
Besides, single- and multi-walled carbon nanotube  can carry a high
current density of the order of $10^9 - 10^{10}$ A/cm$^2$, see e.g.
Refs. \cite{Yao_00,Wei_01,Vajtai_02}. Lastly, metallic CNTs exhibit
a strong, as large as 50-100 times, slowing down of surface
electromagnetic waves \cite{Slepyan_99_PRB,Maksimenko_b04}. Thus, a
combination in CNTs of three key properties,

(i) ballisticity of the electron flow over typical CNT length,

(ii) extremely high current-carrying capacity, and

(iii) strong slowing down of surface electromagnetic waves,

\noindent \noindent allows proposing them as candidates for the
development of nano-sized Chernekov-type emitters -- nano-TWT,
nano-BWO and nano-FEL.

The remainder of the paper is organized as follows. In Sect.
\ref{instability} we derive dispersion equation for electromagnetic
wave coupled with electron beam and discuss its solution in
classical and quantum limits. A solution of the boundary-value
problem for a finite-length CNT is presented in Sect.
\ref{thresholds} allowing evaluation of the absolute instability
gain and the lasing threshold currents. Section \ref{analysis}
contains physical analysis of the results obtained and numerical
estimates for the gain and threshold currents. Concluding remarks
are given in  Sect. \ref{conclusion}.

\section{Radiative instability of electron beam in carbon nanotube}
\label{instability}

\subsection{Self--consistent equation of motion for  electromagnetic wave and electron beam}
\label{diss}

Nanotubes -- quasi-one-dimensional carbon macromolecules -- are
obtained by rolling up of graphene layer into a cylinder. The
transformation can be performed in different manners classified by
the dual index $(n_1,n_2)$. The two integers $n_1$ and $n_2$
represent the vector characterizing the way of turning, with $n_1=0$
for zigzag CNTs, $n_1=n_2$ for armchair CNTs, and $0<n_1\neq n_2$
for chiral CNTs. A nanotube can manifest either metallic or
semiconductor properties, depending on its  radius $R_\mathrm{cn}$
and the direction of rolling up. This correlation arises from the
transverse quantization of charge carrier motion and is due to the
quasi-one-dimensional topology of CNTs; for details see, e.g., Refs.
\cite{Dresselhause_b01,Reich_b04}.

Consider an electron beam moving in an isolated single-walled carbon
nanotube oriented along the $z$-axis. The electron beam can be
injected into the nanotube from  outside by an external source or
can be produced by applying voltage to some section of the nanotube.
Accelerated by the voltage, electrons are injected into the working
region. Independently on the origin of electrons, their motion in
this region is assumed to be ballistic.

As was mentioned in introduction, there is a certain analogy between
a CNT guiding electron beam and macroscopic vacuum electron devices.
The main (and obvious) distinction is the small cross-sectional
radius of CNTs as compared to their macroscopic analogs. As a
result, in CNTs spatial quantization of the electron motion comes
into play and, therefore, classical models of the electron beam
becomes inapplicable. The electron motion in CNTs is governed by
quantum--mechanical equations. In this paper we shall consider the
lasing effect when generated field is rather large, i.e., the
condition
\begin{equation}
E\gg \sqrt{\hbar c}\left(\frac{\omega}{c}\right)^2
\label{classic_condition}
\end{equation}
is fulfilled\cite{Berestetskii}. In this case the electromagnetic
wave has classical character and is described by the classical wave
equation:
\begin{eqnarray}
\nabla \nabla \cdot\mathbf{E}(\mathbf{r},\omega
)-\Delta \mathbf{E}(\mathbf{r},\omega )
=\frac{4\pi i\omega }{c^{2}}\mathbf{j}(\mathbf{r},\omega ).
\label{maxwell}
\end{eqnarray}
If the condition (\ref{classic_condition}) does not hold,  the
number of photons per quantum level becomes too small to apply a
classical approach and the electromagnetic field must be considered
within the quantum electrodynamics. The quantum--electrodynamical
consideration is of importance on the initial stage of the
instability development, when few photons participate in the
process. We leave this stagefor further analysis focusing on the
stage of highly developed instability. Thus, in our model the
electron motion is governed by the Schr\"{o}dinger equation while
the electromagnetic field is described by classical Maxwell
equations. In the right--hand part of the field equation
(\ref{maxwell}) the quantity $\mathbf{j}(\mathbf{r},\omega )$ is the
current density averaged over the quantum states of the electron
beam.

The current density in the working region is defined  by the well--known
equation \cite{dau}:
\begin{eqnarray}
&&\mathbf{j}(\mathbf{r},t)=\frac{e}{2m_e}\left\{
    \psi ^{\ast}(\mathbf{r},t)\mathbf{\hat p}\psi(\mathbf{r},t) -
    (\mathbf{\hat p}\psi ^{\ast }(\mathbf{r},t))\psi(\mathbf{r},t)
    \right\} \cr\rule{0in}{4ex}
&& \qquad\quad-\frac{e^{2}}{m_ec}\left\vert \psi(\mathbf{r},t)
\right\vert^{2}\mathbf{A}(\mathbf{r},t).~~ \label{currr}
\end{eqnarray}
Here $\mathbf{\hat p}=-i\hbar \partial/(\partial \mathbf{r})$ is the
momentum operator and $\mathbf{A}(\mathbf{r},t)$ is the vector
potential of electromagnetic field. Further we neglect the Fermi law
for the electron statistics. This is possible because the number of
excited electrons per quantum level is found to be small even at
superior current densities reachable in
CNTs\cite{Yao_00,Wei_01,Vajtai_02}. Indeed, number of levels in the
interaction volume $V$ is estimated as $\sim Vp^{3}/(2\pi \hbar
)^{3}$, where $p$ is a typical value of quasi--momentum of electrons
in the beam.  The number of electrons in this volume is $\sim n_e
V$, where $n_e$ is the electron density. Then, the number of excited
electrons per level is given by $\eta_e =(2\pi \hbar
)^{3}n_e/p^{3}$. At current density $10^8$ to $10^{10}$ A/cm$^2$ and
an excitation energy of the order of several electronvolts, we find
$\eta_e \sim 10^{-5}-10^{-3}$. Therefore, the exchange interaction
between  electrons in the beam can be neglected.

Let $\psi(\mathbf{r},t=0)=\psi_n(\mathbf{r})$ be the eigenfunction
of an electron  noninteracting with electromagnetic wave and moving
along the CNT. When the interaction is switched on the wavefunction
is represented by the expansion
\begin{equation}
\psi(\mathbf{r},t) =\sum_{l} a_{l}(t)\exp ( -i\varepsilon
_{l}t/\hbar) \psi _{l}(\mathbf{r}) \label{pertubation}
\end{equation}
over a complete set of the unperturbed eigenfunctions  $\psi
_{l}(\mathbf{r})$ with $\varepsilon _{l}$ as  corresponding energy
eigenvalues. For further convenience, we rewrite the coefficients
$a_{l}(t)$ as $a_{l}(t)=\delta_{ln}+\delta a_{l}^{(n)}(t)$, where
$\delta_{ln}$ is the Kronecker symbol. Corrections $\delta
a_{l}^{(n)}(t)$ are due to the electron--electromagnetic field
interaction. Taking into account axial periodicity of the nanotube
potential, the wavefunctions $\psi _{l}(\mathbf{r})$ can be written
in accordance with the Bloch theorem as
\begin{equation}
    \psi _{l}(\mathbf{r})=\exp \left\{ ip_lz/\hbar\right\} \sum\limits_{\tau
}b_{l\tau }\exp \left\{ i\tau z\right\} u_{l\tau
}(\mathbf{r}_{\perp })\label{eigen}.
\end{equation}
Here $p_{l}$  is the axial projection of the quasi--momentum of
$l$-th state, $b_{l\tau}$ are constant coefficient,
 $\tau=2\pi q/a$ are the reciprocal lattice constants,
 $a$ is the CNT spatial period in the axial direction,
$u_{l\tau }(\mathbf{r}_{\perp })$ are functions dependent only on
transverse coordinates, and $q$ are integers. The term $
\sum\limits_{\tau }b_{l\tau }\exp \left\{ i\tau z\right\} u_{l\tau
}(\mathbf{r}_{\perp })$  is periodical in the $z$ direction.

In linear approximation, the contribution to the electron current
(\ref{currr}) originated from the electron--electromagnetic field
interaction  is described by the equation:
\begin{eqnarray}
\delta \mathbf{j}_n(\mathbf{r},t)&=&
    \frac{e}{2m_e}  \sum_l
    \Bigl\{ \delta a_{l}^{(n)\ast }(t)
    \exp [ i( \varepsilon _{l}-\varepsilon_{n}) t/\hbar]
    [ \psi _{l}^{\ast }(\mathbf{r})\mathbf{\hat p}\psi _{n}(\mathbf{r})-(\mathbf{\hat p}\psi _{l}^{\ast }(\mathbf{r}))
     \psi _{n}(\mathbf{r})]
   \cr
&+&
    \delta a_{l}^{(n)}(t)\exp[- i( \varepsilon _{l}-\varepsilon _{n}) t/\hbar]
    [\psi _{n}^{\ast }(\mathbf{r})\mathbf{\hat p}\psi _{l}(\mathbf{r})-(\mathbf{\hat p}\psi _{n}^{\ast}(\mathbf{r}))
    \psi _{l}(\mathbf{r})] \Bigr\} \cr
&-&\frac{e^{2}}{m_ec}\left\vert \psi _{n}\right\vert
^{2}\mathbf{A}(\mathbf{r},t). \label{curadd}
\end{eqnarray}
Then, applying to Schr\"{o}dinger equation standard
perturbation--theory technique \cite{dau}   we obtain the equation
describing the dynamics of the coefficients $\delta a_{l}(t)$:
\begin{eqnarray}
&&i\hbar \sum \frac{\partial \delta a_{l}^{(n)}(t)}{\partial t}\psi
_{l}(\mathbf{r})\exp(-i\varepsilon _{l}t/\hbar)= -\frac{e}{2m_ec}
\cr \rule{0in}{5ex}&&
\quad\times\left[ \mathbf{A}(\mathbf{r},t)\mathbf{\hat p}+\mathbf{\hat
p}\mathbf{A}(\mathbf{r},t)\right] \psi
_{n}(\mathbf{r})\exp(-i\varepsilon _{n}t/\hbar)\,, \label{Shreding}
\end{eqnarray}
which is obtained by substitution of (\ref{pertubation}) into the
Schr\"{o}dinger equation and its subsequent linearization with
respect to the electromagnetic field strength. The  Fourier
transform of Eq. (\ref{Shreding}) gives
\begin{eqnarray}
&&\delta a_{l}^{(n)}\left( \omega \right) =\frac{e}{2m_e\omega \hbar
c}\left\langle l\right\vert \mathbf{A}\Bigl(\mathbf{r},\omega
+\frac{\varepsilon _{l}-\varepsilon
_{n}}{\hbar}\Bigr)\mathbf{\hat p}\nonumber \\
&&\qquad\qquad+\mathbf{\hat p%
}\mathbf{A}\Bigl(\mathbf{r},\omega +\frac{\varepsilon
_{l}-\varepsilon _{n}}{\hbar}\Bigr)\left\vert n\right\rangle \,.
\label{coeffi}
\end{eqnarray}
Here we use the standard ket- and bra- notation of
wavefunctions and matrix elements, $\vert l \rangle = \psi_l
(\mathbf{r})$. Only that terms are preserved in (\ref{coeffi}) which
correspond to resonant interaction between electrons and
electromagnetic field. Contribution of the last term in
(\ref{curadd}) is therefore neglected in (\ref{coeffi}). Performing
the Fourier transform of Eq. (\ref{curadd}) along the axial
coordinate and time, we come to the $k,\omega$--space
interaction--induced current density correction:
\begin{eqnarray}
&&\delta \mathbf{j}_n(k,\mathbf{r}_\perp,\omega) = -
    \frac{e^{2}}{4m_e^{2}c}\sum\limits_{l}B_{nl}(k,\mathbf{r}_\perp,\omega)
    \sum\limits_{\tau ^{\prime }\tau } \cr \rule{0in}{6ex}
&&\quad\times
\left\{-
    \frac{
    b_{l\tau ^{\prime }}^{\ast}b_{n\tau }
    \left[u_{l\tau ^{\prime }}^{\ast }
    \left( \mathbf{\hat p}_{n}+\bm{\tau}\right) +
    \left( \mathbf{\hat p}_{n}+\bm{\tau}\right)
    u_{l\tau ^{\prime }}^{\ast }\right]u_{n\tau }}
    {\hbar \omega +\varepsilon_{l}\left( p_{n}-k\right) -\varepsilon
    _{n}(p_{n})}\right.  \cr \rule{0in}{6ex}
    &&\left.\quad +\frac{b_{n\tau }^{\ast }b_{l\tau^{\prime}}
    \left[u_{n\tau }^{\ast }\left( \mathbf{\hat p}_{n}+\bm{\tau}\right)+
    \left( \mathbf{\hat p}_{n}+\bm{\tau}\right)u_{n\tau }^{\ast }\right]u_{l\tau ^{\prime }}}
    {\hbar \omega +\varepsilon _{n}(p_{n})-\varepsilon _{l}\left(p_{n}+k\right) }\right\}\,.
    \label{fourier}
\end{eqnarray}
For convenience, we have introduced the vector form for the lattice
constant $\tau$: $\bm{\tau}=\tau \mathbf{e}_z$, where $\mathbf{e}_z$
is the unit axial vector. The quasi-momentum operator in matrix
elements is given by $\mathbf{\hat p}_{n}=\{\mathbf{\hat
p}_{\perp},p_n\}$, where axial components $p_{n}$ are $C$-numbers
and  transverse components $\mathbf{\hat p}_{\perp}$ are operators.
These operators act only on the right-adjacent functions. Deriving
(\ref{fourier}), we neglect  the longitudinal component $k$ of the
electromagnetic wave vector in matrix elements since $\hbar k/p_n\ll
1$. Summation over the lattice constants  $\tau$ and $\tau '$ is not
independent: for every $\tau$ in sum,  the value of $\tau '$  must
be such that the values $p_n+\tau-\tau '$ are in the first Brillouin
zone. The coefficients $B_{nl}(k,\mathbf{r}_\perp,\omega)$ are given
by
\begin{eqnarray}
&& B_{nl}(k,\mathbf{r}_\perp,\omega)=\sum\limits_{\tau ^{\prime
}\tau }b_{l\tau ^{\prime }}b_{n\tau }^{\ast } \left\langle u_{n\tau
}\left\vert \left( \mathbf{\hat p}_{n}+\bm{\tau}\right)
\mathbf{A}(k,\mathbf{r}_\perp,\omega )\right. \right. \cr &&
\left.\left.
    \qquad + \mathbf{A}(k,\mathbf{r}_\perp,\omega)\left(
\mathbf{\hat p}_{n}+\bm{\tau}\right) \right\vert u_{l\tau ^{\prime
}}\right\rangle\, . \nonumber
\end{eqnarray}
Substituting then (\ref{fourier}) into (\ref{maxwell}) we come to a
self--consistent field equation necessary for the further analysis.

\subsection{Dispersion equation for electromagnetic wave coupled with electron beam}\label{axcon}

Electromagnetic response properties of an isolated single-walled CNT
was studied in Ref. \onlinecite{Slepyan_99_PRB} on the base of a
tight-binding microscopic model of the CNT conductivity and the
effective boundary conditions for electromagnetic field imposed on
the CNT surface. A detailed analysis of the eigenwave problem has
revealed propagation in CNTs strongly slowed down surface waves
allowing the concept of nanotubes as surface-wave nanowaveguides.
Considering the electron beam as a perturbation, we can use the
dispersion equation for the surface waves and the propagation
constants obtained in Ref. \onlinecite{Slepyan_99_PRB} as a
zero--order approximation. Then, the self--consistent field  of the
electromagnetic wave coupled with electron beam can be presented by
the expansion
\begin{equation}
\mathbf{A}(k,\mathbf{r}_\perp,\omega)=
    \sum\limits_{m}\alpha_{m}(k,\omega)
    \mathbf{A}_{m}(\mathbf{r}_{\perp})\,, \label{expand}
\end{equation}
where vector potentials $\mathbf{A}_{m}(\mathbf{r}_{\perp })$
correspond to the electromagnetic field eigenfunctions evaluated in
Ref. \cite{Slepyan_99_PRB} and $\alpha_{m}(k,\omega )$ are the
coefficients to be found. Substitution of (\ref{expand}),
(\ref{fourier}) and (\ref{curadd}) into (\ref{maxwell}) gives the
system of equations for the electromagnetic field interacting with
the electrons occupying $n$-th state:
\begin{eqnarray}
&&\sum\limits_{m}\left( k^{2}-k_{m}^{2}\right)\alpha_{m}(k,\omega)
    \mathbf{A}_{m}(\mathbf{r}_{\perp })=-\frac{4\pi}{c}
    \frac{e^{2}n_e}{4m_e^{2}c}\sum\limits_{l}B_{nl}(k,\mathbf{r}_\perp,\omega)
\sum\limits_{\tau ^{\prime }\tau }\cr\rule{0in}{6ex} &&\qquad\times
\left\{-\frac{
    b_{l\tau ^{\prime }}^{\ast}b_{n\tau }
    \left[u_{l\tau ^{\prime }}^{\ast }
    \left( \mathbf{\hat p}_{n}+\bm{\tau}\right) +
    \left( \mathbf{\hat p}_{n}+\bm{\tau}\right)
u_{l\tau ^{\prime }}^{\ast }\right]u_{n\tau }}
    {\hbar \omega +\varepsilon_{l}\left( p_{n}-k\right) -\varepsilon
    _{n}(p_{n})}\right.
    \cr\rule{0in}{6ex} &&\left.\qquad
    +\frac{ b_{n\tau }^{\ast}b_{l\tau ^{\prime}}
    \left[u_{n\tau }^{\ast }\left( \mathbf{\hat
p}_{n}+\bm{\tau}\right)+\left( \mathbf{\hat p}_{n}+\bm{\tau}\right) u_{n\tau }^{\ast }\right]
u_{l\tau ^{\prime }}%
}{\hbar \omega +\varepsilon _{n}(p_{n})-\varepsilon _{l}\left(
p_{n}+k\right) }\right\}\,. \label{intermid}
\end{eqnarray}
Here $k_m $ are the wavenumbers corresponding to the physical system
devoid electron beam. As one can see, deriving (\ref{intermid}) we
have proceeded from the single--electron dynamics to the dynamics of
the electron beam: $n_e$ is the electron density. Multiplying left-
and right--hand parts of Eq. (\ref{intermid}) by
$\mathbf{A}_{m}^*(\mathbf{r}_{\perp })$ and  utilizing the
wavefunctions' orthogonality, we come to the dispersion equation as
follows:
\begin{eqnarray}
&&k-k_{m}=- \frac{\omega_L^{2}}{8 k_m m_e c^2}\sum\limits_{l}\vert
B_{nl}^{(m)}\vert ^2 \cr
    &&  \times \left[ \frac{1}{-\hbar \omega
+\varepsilon _{n}(p_{n})-\varepsilon _{l}\left( p_{n}-k\right) }
  +\frac{1}{\hbar \omega +\varepsilon _{n}(p_{n})-\varepsilon
_{l}\left( p_{n}+k\right) }\right]. \label{disperscom}
\end{eqnarray}
The upper index in $B_{nl}^{(m)}$ relates the matrix element with
the corresponding mode of the electromagnetic field
$\mathbf{A}_{m}\left(\mathbf{r}_{\perp }\right)$,
$\omega_L=2\sqrt{\pi e^2 n_e/m_e}$ is the Langmuir frequency of the
electron beam.

The transcendent dispersion equation (\ref{disperscom}) predicts the
existence of a variety of branches of wavenumber $k$.   Among them,
the number of branches to be accounted for is defined by specific
physical parameters of analyzed system. In the vicinity of a
resonance, only terms corresponding to the resonant interaction, one
or several  (in the case of level degeneration), can be kept in the
dispersion equation.  If the difference between levels exceeds the
linewidth, the only the resonant term is of importance.

\subsection{Classical and quantum limits in synchronism conditions}\label{recoil}

Two terms in the right--hand part of Eq. (\ref{disperscom}) dictate two
synchronism conditions  corresponding to the resonant interaction between
electron beam and electromagnetic wave:
\begin{equation}
\pm \hbar \omega +\varepsilon _{n}(p_{n})-\varepsilon _{l}\left(
p_{n}\pm k\right) = 0. \label{resonance}
\end{equation}
The signs ''+'' and ''$-$'' correspond to the absorption and the
emission of photon by electron, respectively. Dependently on the
relation between electron and photon energies, different interaction
regimes are realized. As we restricted ourselves to the case when
the photon momentum is much less than the electron one, the electron
energy $\varepsilon _{l}\left( p_{n}\pm \hbar k\right)$ can be
presented by the truncated Taylor series as
$$
\varepsilon _{l}( p_{n}\pm \hbar k)=\varepsilon _{l}( p_{n})
    \pm  \hbar k\,\frac{\partial \varepsilon_l ( p_{n})}{\partial p_n}
    \equiv \varepsilon _{l}( p_{n})\pm \hbar k\, v_l \, ,
$$
where $v_l$ is the electron group velocity. Then, denominators in
(\ref{disperscom}) can be represented by
\begin{eqnarray}
&&\pm \hbar \omega +\varepsilon _{n}(p_{n})-\varepsilon _{l}(p_{n}\pm k) \nonumber \\
&& \qquad \approx \pm \hbar \left(\omega - k v_l
    \pm \Omega_{nl}\right)+\frac{\hbar^2}{2}\frac{\partial ^{2}\varepsilon
_{l}}{\partial p_{n}^{2}}k^{2}\,. \label{expansion}
\end{eqnarray}
The first  term in the right--hand part of this equation is
analogous to the standard term $\omega - k u \pm \Omega$ in the
synchronism condition\cite{fer}. The only difference is that the
velocity of free electrons is replaced by the group velocity of
quasi--electrons $v_l$ and the undulation frequency is replaced by
the transition frequency $\Omega_{nl}=[\varepsilon
_{n}(p_{n})-\varepsilon _{l}(p_{n})]/\hbar$  between CNT energy
bands. The last term in (\ref{expansion}) originates from the
quantum recoil of electron during emission (absorption) of photon
and induces a red (blue) shift of the transition frequency. This
term is inversely proportional to the electron effective mass
(second derivative of the energy). Let $l=s$ be an electron level
corresponding to the resonant interaction. Then, within the
approximation stated, the dispersion equation takes the form as
follows:
\begin{equation}
k-k_{m}=
    \frac{\displaystyle \frac{2}{\hbar}{b_{ns}^{(m)}}
    \left(\frac{\hbar k^{2}}{2}\frac{\partial ^{2}\varepsilon _{s}}{\partial p_{n}^{2}}-
    \Omega _{ns}\right)}
    {\displaystyle\left( \omega -k v_{s}\right)^{2}-
    \left(
    \frac{\displaystyle\hbar k^{2}}{\displaystyle 2}
    \frac{\displaystyle\partial ^{2}\varepsilon_{s}}{\partial\displaystyle p_{n}^{2}}-\Omega _{ns}
    \right)^{2}}\,, \label{finaldisp}
\end{equation}
where
$$
b_{ns}^{(m)}=-\frac{\omega _{L}^{2}\hbar}{8m_ek_{m}^\prime
c^{2}}\vert B_{ns}^{(m)}\vert^2,~~k_{m}^\prime={\rm Re}(k_m).
$$
In the case of intraband transitions $\Omega _{ns}=0$ and Eq.
(\ref{finaldisp}) takes the form of the dispersion equation for the
instability with the recoil accounted for\cite{Madey}.

Depend on ratio between the  radiation linewidth and the
recoil-induced detuning, two different generation regimes are
realized.  In the low--gain limit \cite{Marshall} the spontaneous
emission linewidth can be estimated as $\Delta \omega/\omega\sim
c/(\omega L)$, where $L$ is the interaction length. If the linewidth
exceeds the recoil energy, the recoil term in the denominator of
(\ref{finaldisp}) can be neglected and the classical interaction
regime is realized. The dispersion equation in that case takes the
traditional form of the second--order Cherenkov resonance:
\begin{equation}
k-k_{m}=
    k^{2}\frac{\partial ^{2}\varepsilon _{s}}{\partial p_{n}^{2}}
    \frac{b_{ns}^{(m)}}{\left( \omega -k v_{s}\right) ^{2}}\,. \label{clasdisp}
\end{equation}
The  \emph{spatial increment} of the instability $k''={\rm Im}(k)$
 can be estimated using the method of weakly
coupled modes\cite{kinetics}.  According to this method, interaction
between the electromagnetic wave and the electron beam is essential
only in the vicinity of the point $(\omega_0, k_0=\omega_0/v_s)$
where the  dispersion curves of noninteracting modes, $\omega -k
v_{s} = 0$ and $k(\omega)=k_m(\omega)$, are crossed. Then $k_m$ is
represented by the expansion
\begin{equation}
 k_m(\omega)=k_0+\left.\frac{\partial k_m(\omega)}{\partial \omega}\right|_{\omega=\omega_0} (\omega
- \omega_0)\,.  \label{represent}
\end{equation}
Substitution of this expansion and $k=k_0+\Delta k$  into
(\ref{clasdisp}) results in a third--order algebraic equation with
respect to $\Delta k$. From this equation, the instability spatial
increment  is estimated at the frequency $\omega=\omega_0$ as
\begin{equation}
\left\vert \Delta k''\right\vert =\frac{\sqrt{3}}{2}\left\vert
b_{nn}^{(m)}\frac{\partial ^{2}\varepsilon _{n}}{\partial
p_{n}^{2}}\frac{k^{2}}{v_{n}^{2}}\right\vert ^{1/3}\,,
\label{increment}
\end{equation}
where $ \Delta k''={\rm Im}(\Delta k)$. Since $b_{nn}\sim n_e$, the
increment is found to be the 3-rd root of the electron density. Such
a dependence is typical for the Compton--type radiative instability
\cite{Marshall}.

In the opposite case, when the linewidth is less then the difference
between the emission and the absorption frequencies, we fall into
regime of the \emph{strong quantum recoil impact}. In this case,
only the term corresponding to the emission survives in the
dispersion equation (\ref{disperscom}), which therefore is reduced
to
\begin{equation}
k-k_{m}=\frac{b_{nn}^{(m)}}{\hbar}\frac{1}{%
 \omega -v_{s}k- \frac{\displaystyle\hbar
}{\displaystyle 2}\frac{\displaystyle\partial ^{2}\varepsilon
_{n}}{\displaystyle \partial p_{n}^{2}}k^{2}}. \label{dispquant}
\end{equation}
As a result, the instability increment is given by
\begin{equation}
\left\vert \Delta k''\right\vert =\left\vert \frac{b_{nn}^{(m)}}{\hbar v_{n}}%
\right\vert ^{1/2}\,,  \label{incrementq}
\end{equation}
i.e., turns out to be proportional to the square root of the
electron density.

Below we present a detail discussion of the different generation
regimes and give some numerical estimates of physical parameters
corresponding to these regimes.

\section{Starting currents and their dependence on the nanotube
length}\label{thresholds}

\subsection{Boundary conditions for a finite--length nanotube}\label{bound}

In sections \ref{axcon} and \ref{recoil}, dispersion equations have
been derived providing us with  wavenumber eigenvalues in an
infinite--length CNT guiding  electron beam. As a next step, the
system must be imposed by edge conditions accounting for the finite
length of the interaction zone. These conditions are stated as the
requirement to perturbations of the electron and current densities,
generated by the electron beam -- electromagnetic wave interaction,
to be zero at the input of the working zone, i.e.
\begin{equation}
\delta n_e (z=0) =\delta j_n (z=0) =0\,. \label{resonator0}
\end{equation}
The condition that the tangential electric field component and the
axial component of the magnetic field be continuous on the CNT
surface yields additional boundary condition. We write it in the
simplified form \cite{zvelto} as
\begin{equation}
E(z=0)=\alpha E(z=L)\,, \label{resonator}
\end{equation}
where $\alpha$ is  the reflection coefficient of electromagnetic
field from the working zone boundaries.

The field distribution in a finite-length system consisting of
several parts can be found by solving electrodynamical problem in
each region separately and then joining the solutions by means of
boundary conditions. In the interaction region, the electromagnetic
field is given by
\begin{equation}
E(z)\sim \sum_{i=1}^{N}c_i \exp \bigl(ik^{(i)} z\bigr)\,,
\label{field3}
\end{equation}
where the summation is performed over all electromagnetic modes in
CNT; the wavenumbers $k^{(i)}$ are determined by corresponding
dispersion equations. Note that the reflection of electromagnetic
waves from boundaries back into the working zone creates positive
feedback in the system and thus allows accumulation of the
electromagnetic energy and provides an oscillator regime.

\subsection{Starting current at a large quantum recoil}\label{startrecoil}

In the quantum interaction regime, when the quantum recoil exceeds
the  linewidth, the instability is described by the quadratic
dispersion equation (\ref{dispquant}) with solutions $k^{(1)}$ and
$k^{(2)}$. Consequently, the electric field and the perturbation of
the current density in the working zone are given by
\begin{eqnarray}
&&E\sim c_{1}\exp \bigl( ik^{(1)}z\bigr) +c_{2}\exp \bigl(
ik^{(2)}z\bigr)\, , \label{fieldq_E}\\ \rule{0in}{4ex}
&&\delta j_n\sim \frac{c_{1}}{\delta _{1}}\exp \bigl( ik^{(1)}z\bigr) +\frac{c_{2}}{%
\delta _{2}}\exp \bigl( ik^{(2)}z\bigr)\,.
\label{fieldq}
\end{eqnarray}
The coefficients
\begin{equation}
\delta _{1,2}=1 -\frac{v_{n}}{\omega}
k^{(1,2)}+\frac{\hbar}{2\omega}\, \frac{\partial
^{2}\varepsilon_{n}} {\partial p_{n}^{2}}\,k^{(1,2)2} \label{deviation}
\end{equation}
introduce deviations of  the wavenumbers  $k^{(1)}$ and $k^{(2)}$
from the synchronism, and the coefficients $c_{i}$ are determined
from the boundary conditions as was discussed in Sect. \ref{bound}.
Using  the boundary condition (\ref{resonator0}) and
(\ref{resonator}),  we arrive at the linear system for $c_{i}$ as
follows:
\begin{eqnarray}
\begin{array}{l}
c_{1}+c_{2}=\alpha \bigl[ c_{1}\exp \bigl( ik^{(1)}L\bigr)
+c_{2}\exp \bigl( ik^{(2)}L\bigr) \bigr]\,, \\
\rule{0in}{4ex}
{\displaystyle \frac{c_{1}}{\delta
_{1}}+\frac{c_{2}}{\delta _{2}}=0\,. }\label{thresh_syst}
\end{array}
\end{eqnarray}
Nontrivial solution of this system is determined by the equation
\begin{eqnarray}
\delta _{1}\bigl[ 1-\alpha \exp \bigl( i k^{(1)} L\bigr)
\bigr] -\delta _{2}\bigl[ 1-\alpha \exp \bigl( i k^{(2)} L\bigr)
\bigr]=0\,.
 \label{generation}
\end{eqnarray}
A current density satisfying equation (\ref{generation}) is the
\emph{threshold current density} of the generation. To evaluate this
quantity, characteristic equation (\ref{generation}) must be solved
together with equation (\ref{dispquant}). Substituting  the roots
\begin{eqnarray}
k^{(1,2)}=k_{m,\mathrm{ch}}+\frac{b_{nn}^{(m)}}{ \hbar
v_n(k_\mathrm{ch}-k^\prime_{m}) }  \label{roots}
\end{eqnarray}
of the dispersion equation (\ref{dispquant}), with $k_\mathrm{ch}$
extracted from the synchronism condition $\omega
-k_\mathrm{ch}v_{n}+(\hbar k_\mathrm{ch}^{2}/2)\partial
^{2}\varepsilon _{n}/
\partial p_{n}^{2}=0$, into (\ref{generation}) and solving the resulting
equation with respect to the current density, we obtain
\begin{equation}
\frac{b_{nn}^{(m)}}{\hbar v_n} L^{2}\frac{\sin ^{2}x
}{ x^{2}}=1-\left\vert \alpha \right\vert +Lk_{m}''\,,
\label{threshold}
\end{equation}
where \begin{equation}
\label{dev}
x=\left(\omega -k^\prime_{m}v_{n}+\frac{\hbar k^{\prime
2}_{m}}{2}\frac{\partial ^{2}\varepsilon _{n}}{
\partial p_{n}^{2}}\right)\frac{L}{2c}
\end{equation}
is the dimensionless off--synchronism parameter.

Physically, Eq. (\ref{threshold}) establishes the energy balance in
the working zone. Its  left--hand  part determines the radiation
production which is therefore proportional to the electron density
$n_e$ and to the squared interaction length. The factor $\sin^{2} x
/ x^{2}$ determines the so called \emph{gain curve} --- the gain
dependence  on the off-synchronism parameter $x$. In the case
considered the gain curve is symmetrical with respect to $x=0$ and
is maximal at zero deviation $x$. Further we compare this result
with the classical case of small recoil and demonstrate significant
difference in the behavior of gain curves. The term $1-\alpha$ in
the right--hand part of (\ref{threshold}) corresponds to the
radiation leakage through the boundaries of the interaction zone
while the last term specifies the radiation absorption by nanotube.

The energy balance equation (\ref{threshold}) allows the evaluation
of the threshold current density. If the current density in CNT
exceeds the threshold value,  the generation process is developed.
The characteristic time  of the instability development is inversely
proportional to the \emph{absolute instability increment}
$\omega''={\rm Im}(\omega)$, which is derived by solving the
generation equation (\ref{generation}) with respect to $\omega (k)$.
In the lowgain regime \cite{Marshall}, which implies  the conditions
$\vert \Delta k^{\prime \prime} \vert L \ll 1$ and $1- \alpha \ll 1$
to be fulfilled, the increment is given by:
\begin{eqnarray}
\omega_m ^{\prime \prime }=\left[\frac{\partial k_{m}}{\partial
\omega }\right]^{-1}
    \left( \frac{b_{nn}^{(m)}}{\hbar v_n} L\frac{\sin ^{2}x }{x^{2}}-\frac{1-\left\vert
\alpha \right\vert }{L}
 - k_{m}''\right).
\label{increm_increm}
\end{eqnarray}
In the linear stage of the radiative instability development, the
electromagnetic field grows with time as $\exp(\omega''_m t)$.

\subsection{Starting current in the classical regime of interaction} \label{startclass}

\begin{figure}[htb]
\centering {
\includegraphics[width=3.5in]{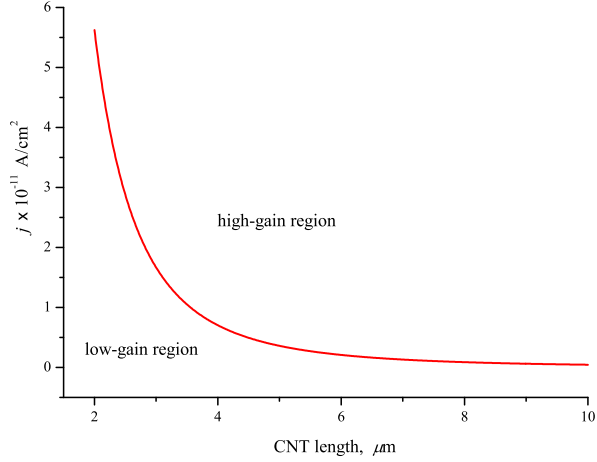}
\caption{ Demarcation between low- and high- gain regimes  of the
generation equation \eqref{thresh_eq_cl}.}\label{gai} }
\end{figure}

In the case when quantum recoil can be neglected, the dispersion
equation (\ref{clasdisp}) gives three roots
\begin{eqnarray}
k^{(1)}=k_{m}-b_{nn}^{(m)}\frac{\partial ^{2}\varepsilon
_{n}}{\partial p_{n}^{2}}\frac{k^{\prime 2}_m}{\left( \omega
-v_{n}k^\prime_{m}\right) ^{2}} \nonumber \\\rule{0in}{5ex}
k^{(2,3)}=k_\mathrm{ch}\pm \frac{i}{v_n}\sqrt{b_{nn}^{(m)}%
\frac{\partial ^{2}\varepsilon _{n}}{\partial p_{n}^{2}}\frac{k^{\prime 2}_m}{%
k_\mathrm{ch}-k^\prime_{m}}}.%
\label{class_root}
\end{eqnarray}
and, consequently, electromagnetic field in the interaction  region
is given by Eq. (\ref{field3}) with $N=3$. Correspondingly,
perturbations of the electron and the current densities in the beam
are written as
\begin{eqnarray}
\delta j_n\sim \sum\limits_{i=1}^3 \frac{c_{i}}{\nu_{i}^{2}}\,,
    \qquad \delta j_n-v_{n}\delta n_e\sim
\sum\limits_{i=1}^{3}\frac{c_{i}}{\nu_{i}}\,,%
\label{densities}
\end{eqnarray}
where deviations $\nu _{i}$  are given by (\ref{deviation}) with the
last term omitted, i.e.,  $\nu _{i}=1 -k^{(i)}v_{n}/\omega $. Then,
by analogy with the previous section, we obtain the linear system
\begin{eqnarray}
\begin{array}{l}
c_{1}+c_{2}+c_{3}=\alpha \left[ c_{1}\exp \bigl( ik^{(1)}L\bigr)
    \right. \\ \rule{0in}{4ex}\left.\qquad
    +c_{2}\exp \bigl( ik^{(2)}L\bigr)+ c_{3}\exp \bigl( ik^{(3)}L\bigr) \right]\,,
 \\
\rule{0in}{4ex} \displaystyle \frac{c_{1}}{\nu
_{1}}+\frac{c_{2}}{\nu _{2}}+\frac{c_{3}}{\nu _{3}}=0\,,
\\ \rule{0in}{4ex}\displaystyle
\frac{c_{1}}{\nu _{1}^{2}}+\frac{c_{2}}{\nu _{2}^{2}}+\frac{c_{3}}{%
\nu _{3}^{2}}=0\,,
\end{array}
\label{thresh_syst_classic}
\end{eqnarray}
and  corresponding generation equation
\begin{eqnarray}
&& \nu _{1}^{2}\left( \nu _{2}-\nu _{3}\right) \left[ 1-\alpha
\exp
\bigl( ik^{(1)}L\bigr) \right] \nonumber  \\
&& \quad- \nu _{2}^{2}( \nu _{1}-\nu_{3})
    \left[ 1-\alpha \exp \bigl( ik^{(2)}L\bigr) \right] \nonumber \\
&& \quad+\nu _{3}^{2}( \nu _{1}-\nu _{2}) \left[ 1-\alpha
    \exp \bigl( ik^{(3)}L\big) \right] =0\, .\label{thresh_eq_cl}
\end{eqnarray}
This equation we solve in the low-gain limit, which is determined by
the condition $k''_z L\le 1$. The curve depicted in Fig. \ref{gai}
divides out  areas of parameters corresponding to low- and high-gain
regimes, respectively.
Then, solutions  of Eq. (\ref{thresh_eq_cl}) -- the threshold
current and the temporal instability increment -- are given by
\begin{eqnarray}
&&\frac{b_{nn}^{(m)}}{v_n^2} \frac{\partial
^{2}\varepsilon_{n}}{\partial p_{n}^{2}}kL^3\,í
    \frac{x\cos x-\sin x}{x^{3}}=
    1-\left\vert \alpha \right\vert
+Lk_{m}''\,, ~~~ \label{thresh_thresh_classic0} \\ \rule{0in}{5ex}
&&\omega_m ^{\prime \prime }=\left[\frac{\partial k_{m}}{\partial
\omega }\right]^{-1}\left[\frac{b_{nn}^{(m)}}{v_n^2}
\frac{\partial ^{2}\varepsilon _{n}}{\partial
p_{n}^{2}}L^{2}\,\frac{x\cos x-\sin
x}{x^{3}} \right. \nonumber \\ \rule{0in}{4ex}
&&\qquad\left. -\frac{1-\left\vert \alpha \right\vert }{L}+k_{m}''\right]\,. \label{incr_incr_classic}
\end{eqnarray}
with the parameter $x$ defined by Eq. (\ref{dev}). As follows from
the balance equations (\ref{threshold}) and
(\ref{thresh_thresh_classic0}), in the quantum interaction regime
the radiation production per unit length is characterized by the
linear dependence on $L$, while this dependence becomes quadratic in
the classical regime. Besides, the gain curves display  distinctive
behavior in these two cases. As different from the quantum
interaction regime,  in the classical limit the gain curve has
asymmetrical character\cite{Marshall}  due to the interference of
absorption and emission processes separated in this case by a
frequency gap narrower then the linewidth. As a result, sign of the
absolute instability increment depends on the sign of the
synchronism detuning. At positive detuning the system is closer to
the absorption frequency while negative detuning moves the system to
the emission frequency.
\begin{figure}[htb]
\centering {
{\includegraphics[width=3.5in]{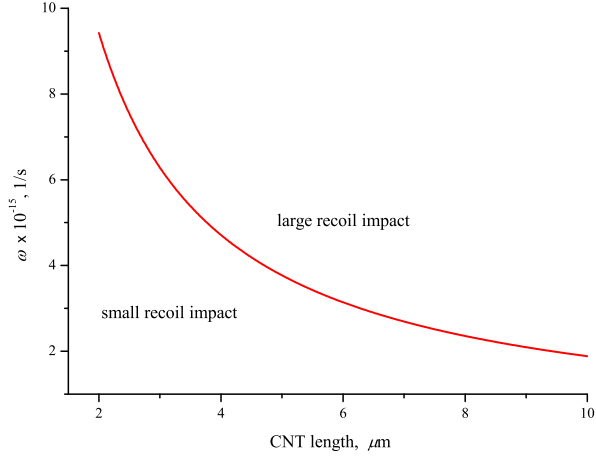}}\\
\caption{ The curve divides the regions of parameters with small and
large impact of the quantum recoil on the generation.  The curve has
been obtained for the low-gain regime. }\label{rec} }
\end{figure}

Qualitatively,  the classical and quantum interaction regimes  are
divided by the demarcation line depicted in Fig. \ref{rec}. In the
area above the line the quantum recoil at the generation must be
taken into account while in the area below the line this effect can
be ignored. The line course can easily be explained just by the
increase of the photon energy with frequency. In addition, the
increase of the generation length $L$ leads to narrowing of the gain
line and, as a result, the quantum recoil comes into play at smaller
frequencies.

\subsection{The role of  electron spread}\label{spread}

If electrons in the beam are distributed over a large number of
energy levels and energy spread significantly exceeds the gap
between emission and absorption lines,  the total current is
obtained by summation over this distribution. The generalization of
Eq. (\ref{clasdisp}) on this case is obvious:
\begin{equation}
k-k_{m}(\omega )=-b_{nn}^{(m)}\int dvf(v)%
\frac{\partial ^{2}\varepsilon_n }{\partial
p_n^{2}}\frac{k^{2}}{\left( \omega -vk\right) ^{2}}\,. \nonumber
\end{equation}
This equation can be rewritten in the form, conventional in plasma
physics\cite{dau}:
\begin{equation}
k-k_{m}(\omega)=b_{nn}^{(m)}k\,\frac{\partial^{2}\varepsilon_n
}{\partial p_n^{2}} \int \frac{\partial f(v)}{
\partial v}\frac{dv}{\omega -vk+i0}\,.  \label{distribution}
\end{equation}
Deriving (\ref{distribution}) we assumed the dependence of the
normalized distribution function ($\int f(v) dv = 1$) on the group
velocity to be narrower than corresponding dependences of the matrix
element $b_{nn}^{(m)}$ and  second derivative of the energy
$\partial ^{2}\varepsilon_n/\partial p_n^{2}$. Then, considering the
group velocity spread exceeding the spontaneous emission linewidth,
$\omega \Delta v/c \gg c/(\omega L)$,  in (\ref{distribution}) we
can make use the standard representation
\begin{equation}
 \frac{1}{\omega -vk+i0}={\cal P}\frac{1}{\omega -vk}-i\pi \delta
\left( \omega -vk\right)\,. \label{landau}
\end{equation}
The principal value of the integral determines  the real-valued
component which is out of our interest.

If resonant interaction between electron beam and electromagnetic
field occurs in the region of the negative derivative of the
distribution function, i.e.  $\partial f(v)/\partial v<0$, then
$k''>0$ and the generation process is not developed (we choose the
$\exp(i k L) $ dependence). This is because the majority of
electrons in that case have velocities smaller then the resonant
velocity and therefore they absorb the electromagnetic wave energy.
Such a situation takes place in equilibrium, when the number of
particles occupying energy level grows less with the level energy
increase. In such a system, an initial perturbation attenuates. This
process is commonly known as the Landau attenuation.

If the resonance is in the region with positive derivative
$\partial f(v)/\partial v>0$, the radiative instability is
possible and obeys the condition
\begin{equation}
k_{m}^{\prime \prime }-\pi b_{nn}^{(m)}\frac{\partial ^{2}\varepsilon _{n}}{%
\partial p_{n}^{2}}\left. \frac{\partial f(v)}{\partial v}\right\vert
_{v=\omega /k}\,<0\,,  \label{hotdisp}
\end{equation}
which originates from the requirement $k'' < 0$  and from Eqs.
(\ref{distribution}) and (\ref{landau}). The condition
(\ref{hotdisp}) expresses the  excess of emission over absorption.
As one can see, the emission per unit length does not depend on the
interaction length.

\begin{figure}[htb]
\centering {
{\includegraphics[width=3.5in]{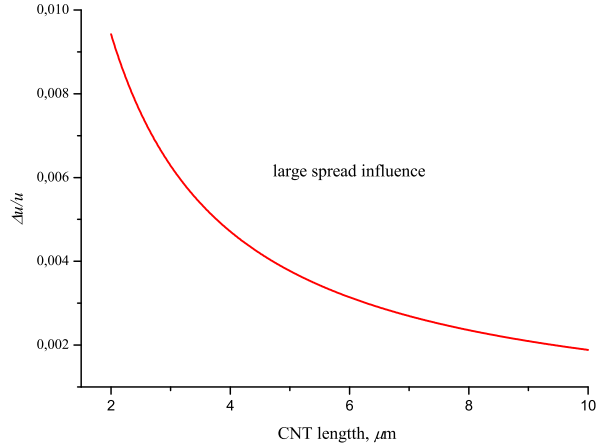}}\\
\caption{ The regions of parameters with small (below curve) and strong (above the curve) influence of the electron beam spread.  }\label{spr} }
\end{figure}

The imaginary part of the wavenumber $k$ describes the asymptotic
exponential behavior of the electromagnetic field in a continuous
medium. To reach generation in a finite region, a corresponding
boundary conditions must be imposed. At a large spread, when the
resonant term in (\ref{distribution}) can be presented by
(\ref{landau}), the dispersion equation (\ref{distribution}) has the
only root. Using (\ref{resonator}) we arrive at the relation $
c_1=\alpha \exp \big( ik^{(1)}L\big) c_1\,,  \label{boundhot} $
which dictates the generation equation as
\begin{equation}
1-\alpha \exp \big( ik^{(1)}L\big)=0.  \label{generhot}
\end{equation}
For the Cherenkov radiation mechanism, solution of  (\ref{generhot})
leads to the equations as follows for the threshold current density
and the absolute instability increment:
\begin{eqnarray}
&& \pi b_{nn}^{(m)} \frac{\partial ^{2}\varepsilon_n }{\partial
p_n^{2}}L\left. \frac{\partial f(v)}{\partial v}\right\vert
_{v=\omega /k}=1-\left\vert \alpha \right\vert + Lk_{m}''\ \label{hotsolv0}\,,~~~ \\ \rule{0in}{5ex}
&& \omega_m ^{\prime \prime }=
    \left[\frac{\partial k^\prime_{m}}{\partial \omega }\right]^{-1}
    \left[\pi b_{nn}^{(m)}\frac{\partial
^{2}\varepsilon_n }{\partial p^{2}}\left. \frac{\partial
f(v)}{\partial v}\right\vert
_{v=\omega /k} \right. \nonumber \\ \rule{0in}{5ex}
&& \qquad\left. -\frac{1-\left\vert \alpha \right\vert }{L}-k''_{m}\right]\,. \label{hotsolv}
\end{eqnarray}
Equation (\ref{hotsolv}) shows that the production of stimulated
radiation in the case of large spread is defined by the spread and
falls down with its increase. The line dividing the range of
parameters into two domains, with weak and strong influence of the
energy spread, is depicted in Fig. \ref{spr}. With the CNT length
increase the role of the spread also rises due to the gain line
narrowing.

The extension of the obtained generation conditions  to the case of
interband transitions (i.e., to the undulator regime)  is obvious
and, in accordance with the dispersion equation (\ref{finaldisp})
for the undulator regime, is achieved by the substitution $\omega
-kv_{n}\to\omega-kv_{s}-\Omega_{ns}$ in the off--synchronism
parameter (\ref{dev}), and the substitution $\hbar k^{2} \partial
^{2}\varepsilon _{n}/\partial p_{n}^{2}\to\hbar k^{2} \partial
^{2}\varepsilon _{s}/\partial p_{n}^{2}-2\Omega _{ns}$ in
expressions for the threshold current (\ref{threshold}),
(\ref{thresh_thresh_classic0}) and (\ref{hotsolv0}),  and for the
absolute instability increment (\ref{increm_increm}),
(\ref{incr_incr_classic}) and (\ref{hotsolv}).

The analytics presented in this section implies fulfillment of
several simplifying approximations: smallness of the photon momentum
$\hbar k/p\ll1$, small or large influence of the electron recoil on
the emission (absorption), small or large electron spread).
Obviously, the analytical approaches do not work in intermediate
cases; equation (\ref{disperscom}) supplemented by corresponding
boundary conditions requires numerical integration. The number of
roots of the dispersion equation to be accounted for and
corresponding number of boundary conditions to be imposed is
dictated by concrete physical parameters of the system being
considered.

\section{Physical analysis and numerical estimates}
\label{analysis}

In Sect. \ref{diss} it has been stated that classical treatment of
electromagnetic field is valid if the field strength $E_e$ amounts
to a certain sufficiently large value. This value is determined by
the condition imposed on the number of photons per energy level to
exceed unity\cite{dau}. At the initial stage of the instability
development, with less then one photon per energy level, the photon
dynamics is described within the quantum electrodynamics formalism.

Usually, the number of photons per energy level is given by $n_{ph}
\left(c/\omega\right)^3$, where $n_{ph}$ is the photon number per
unit volume while the quantity $(\omega/c)^3$ determines the number
of photon levels lying below the energy $\hbar\omega$. As different
from that, in the case of high--coherent laser radiation the
radiation is concentrated in a narrow spectral range $\Delta \omega
\sim c/L$. As a result, the parameter defining the possibility of
classical consideration of electromagnetic waves -- the number of
photons per energy level -- is derived as the density of the beam's
kinetic energy converted to electromagnetic field divided by the
photon energy and the number of levels below $\hbar\omega$. The
ratio is found to be
$$
\sim \eta_{ph}\frac{j}{ve}\left( \frac{c}{\omega }\right)
^{2}L\frac{k_m c }{\omega}\frac{m c^2 (\gamma-1)}{\hbar\omega}\,,
$$
where $j$ is the current density, $v$ is the  electron velocity,
$\eta_{ph}$ is the efficiency of the transfer  of electron kinetic
energy to electromagnetic  field. For infrared photons and electrons
of several electron-Volt energy and $\sim10$ $\mu$m length nanotube,
the photon number per energy level exceeds unity (i.e., the
classical treatment is possible) if $\eta_{ph}>10^{-5}$. Since the
initial stage of the instability development is beyond the scope of
our paper, the parameter $\eta_{ph}$ can be estimated from the
relation $\eta_{ph}\sim 1/(k L)\sim 0.02$, which corresponds to
so-called nonlinear saturation regime \cite{Marshall} and determines
the electron beam energy conversion in saturation. Therefore,
generation threshold and nonlinear stage of the instability
development can be considered classically.

\begin{figure}[th]
\begin{center}
\includegraphics*[width=3.3 in]{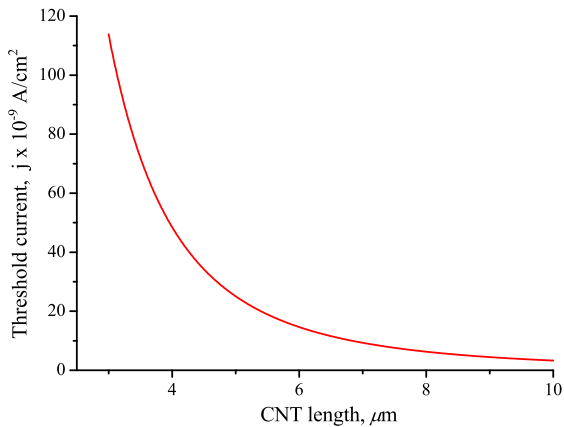}
\end{center}
\caption{ The dependence of threshold current density on nanotube
length.}\label{fig1}
\end{figure}

\begin{figure}[htb]
\centering {
{\includegraphics[width=3.2in]{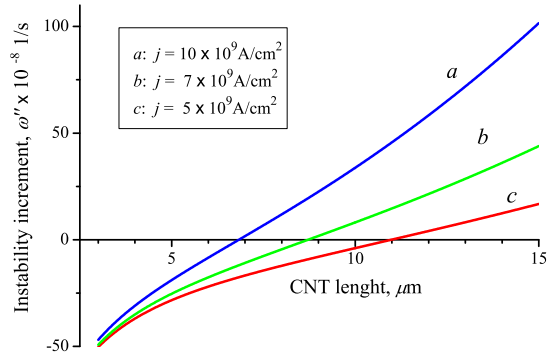}}\\
\caption{ Instability increment \textit{vs} nanotube length at different electron current densities.
}\label{fig2} }
\end{figure}

A simplest way to realize nanoFEL in carbon nanotube is to inject
into it a high energy external electron beam. Since the velocity of
free electron is  $v(\mathrm{cm/s})=5.7 \times 10^{7}\sqrt{\varphi
({\rm eV})}$, in order to accelerate electrons up to velocities
providing the synchronism regime (with 50-100 times wave slowing
down predicted in Ref. \onlinecite{Slepyan_99_PRB}), it is necessary
to apply voltage of $\varphi\sim 7$ eV. If the CNT diameter is such
that its product with the electron transversal momentum is $p_\perp
D /\hbar \sim 10 - 100$, the electron motion can be treated as
classical. In that case, the term in the right part of the
dispersion equation (\ref{clasdisp}) can be modified in the
following way
\begin{equation}
b_{nn}^{(m)}\frac{\partial ^{2}\varepsilon
_{n}}{\partial p_{n}^{2}}\frac{k^{2}}{\left( \omega -v_{n}k\right) ^{2}}%
\sim \omega _{L}^{2}\frac{\left( \mathbf{v}\mathbf{e}\right) ^{2}}{2k_{m}^\prime c^{2}}%
\frac{k^{2}}{\left( \omega -v k\right) ^{2}} . \label{similarity}
\end{equation}
where $\mathbf{v}$ is the classical electron velocity and
$\mathbf{e}$ is polarization vector for the electromagnetic mode
considered. This simplification, after substitution of
\eqref{similarity} into Eqs. \eqref{thresh_thresh_classic0} and
\eqref{incr_incr_classic}, allows us estimate the threshold current
required  to start the generation process  and the instability
increment, respectively. The dependences of these quantities on the
CNT length are depicted in Figs. \ref{fig1} and \ref{fig2}.
Calculations have been done for 1$~\mu$m radiation wavelength and
for the reflection coefficient from the working zone boundaries
$\alpha=0.99$. Generation in the terahertz range would require
higher current density.
It follows from Fig. \ref{fig2} that the gain for CNT is extremely
large as compared with macroscopic electronic devices. For chosen
parameters, the generation development  starts when the CNT length
is about 6 microns or larger, what is technologically routine range.
Therefore, our calculations demonstrate that the development of
CNT-based nanoFEL is already possible at the current stage of
nanotechnology. The characteristic time of the instability evolution
is inversely proportional to the instability increment and for $10\,
\mu {\rm m}$ nanotube  is a fraction of nanosecond.

A positive feedback is required for the realization of oscillator
regime; reflection from CNT ends \cite{Slepyan_06_PRB} can serve as
a possible mechanism of the feedback. The reflection can by
intensified by variation of the CNT generic parameters, proper
selection of surrounding medium and using other methods commonly
applied in laser physics and electronics. An alternative mechanism
providing the feedback is excitation of backward modes propagating
oppositely to electron flow. The backward modes are possible because
CNTs are periodic along their axis and, consequently, their
eigenmodes are Bloch modes containing waves with both positive and
negative phase velocities. As a result, there exist Bloch modes with
group velocity  directed oppositely to the electron velocity -- the
backward modes \cite{new}. One of the waves of the backward mode
having a positive phase velocity can be synchronized with the
electron flow. In this case the positive feedback is provided
automatically.

The instability  process is developed only if the electron free-path
length is comparable or even exceeds the working zone length, i.e.,
the electron motion is ballistic  within the zone. Otherwise, random
collisions of electrons cause a phase shift which prevents the
electron flow bunching  and brakes the radiation coherence. As was
mentioned above, in metallic single-walled CNTs the free-path length
is about several microns \cite{Frank_98,Berger_02,Berger_03}. A
longer ballisticity area can be provided by proper external
conditions. For example, in a regular array of oriented nanotubes
the suppression of electron collisions with atoms can be achieved
using the properties of the electron diffraction in periodical
structures. In a densely packed array of CNTs --- CNT bundle ---
nanotubes form a lattice with the distance between CNTs' axes
$2R_\mathrm{cn}+d$, where $d\approx 3.2\,$\AA~ is the interlayer
distance in graphite. Correspondingly, the reciprocal lattice vector
in such a lattice has the value ${h}=2\pi/(2R_\mathrm{cn}+d)$. From
the principle of uncertainty we can estimate the transverse
component of the momentum by $p_\bot/\hbar\sim 2\pi /2R_{cn}$.
Obviously, the Bragg condition $|\mathbf{p}_\bot + \bm{h}|\approx
|\mathbf{p}_\bot|$ can be fulfilled for a large portion of electrons
passing the bundle and six-wave diffraction\cite{multi} can be
realized. Owing to the diffraction, electrons are concentrated in
domains free of atoms and, therefore, scattering is weak for such
electrons. Analogous situation meet in the Bormann effect
\cite{multi} for hard X-rays passing through a crystal. Owing to
this effect, a significant increase of the photon free path is
observed.

Even if generation conditions are provided by the use of external
electron beam,  the idea to exploit  intrinsic  electrons of CNTs
looks very attractive because it would solve the dramatic problem to
focus an external electron beam into a spot of the CNT diameter
size. Typical velocity of $\pi$-electrons excited to energy of
several electronvolts is about\cite{Reich_b04} $10^8$ cm/s. For such
electrons, the synchronism condition requires the electromagnetic
wave slowing down as large as $300$ times, which is much larger than
the theoretical estimate\cite{Slepyan_99_PRB} gives for CNTs.

In such a situation, special configurations providing higher group
velocity are extremely desirable; otherwise, stronger excitation of
electrons is necessary to fulfil the Cherenkov generation condition.
Fortunately, as compared to vacuum electronic devices, stimulated
emission in CNTs features a set of new promising properties. In
macroscopic Cherenkov FELs the electron energy ordinarily rises with
the electron velocity and, in nonrelativistic regime, quadratically
depends on the momentum (and velocity). As a result, the only way to
reach the synchronism condition in that case is to increase the
electron beam energy. For a collective (quasi-) electrons in CNT
such is not the case. Indeed, the electron group velocity, which is
analog of the velocity for quasi-particles, is determined by the
properties of the whole system and may demonstrate nontrivial
dependence on the quasi-momentum. Locally, the quasi-particle
velocity may recede as energy rises. Correspondingly, local maxima
of the group velocity may appear.  If one seeks the synchronism
condition  for a low-energy quasi-particle, it is advantageous to
choose parameters in the vicinity of the group velocity local
maxima. It allows attaining the synchronism in a relatively low
accelerating potential and, therefore, significantly reduces the CNT
energy load.

Let us exemplify the statement considering an isolated straight
$(q,q)$ armchair CNT. The dispersion law of $\pi$-electrons in such
a CNT is given by\cite{Dresselhause_b01}
\begin{eqnarray}
\varepsilon_l (p)&=&\mp \gamma _{0}\left[1\pm 4\cos(
    {\pi l}/{q})
    \cos({a}p) \right. \cr
\rule{0in}{4ex}
 &&
    \left.+4\cos^{2}({a}p)\right]^{1/2},
 \label{RabA}
\end{eqnarray}
where $\gamma_{0} \simeq 2.7$ eV is the overlap integral,
$l=1,\dots,2q$, $a=\sqrt{3}b/2\hbar,\, b=1.42\, \textrm{\AA}$ is the
interatomic distance in graphite. The upper and lower signs refer to
the conduction and valence bands, respectively. The group velocity
corresponding to this law is
 \begin{eqnarray}
&&v_{l}= \mp 2\gamma_0 {a}\sin \left({pa}\right)  \cr\rule{0in}{5ex}
    && ~~\times \frac{\displaystyle \mp\cos \left( {\pi l }/{q}\right)
    -2\cos \left( {pa}\right) }
    { \displaystyle
    \left[ 1\pm 4\cos \left(\pi l/{q}\right) \cos \left( {pa} \right)
    +4\cos ^{2}\left( {pa}\right) \right] ^{1/2}}\,.\,\,\, \label{grvel}
\end{eqnarray}
\begin{figure}[htb]
 \includegraphics[width=3.1in]{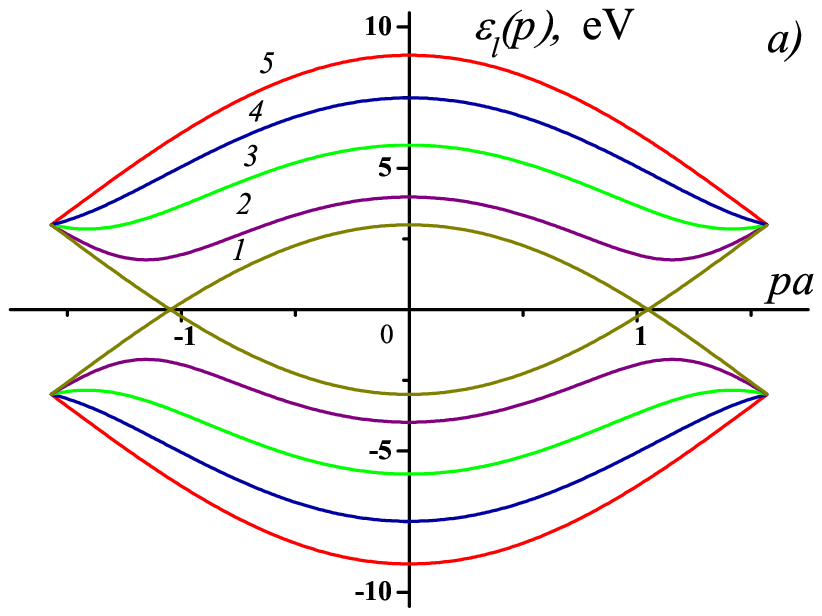}
\includegraphics[width=3.1in]{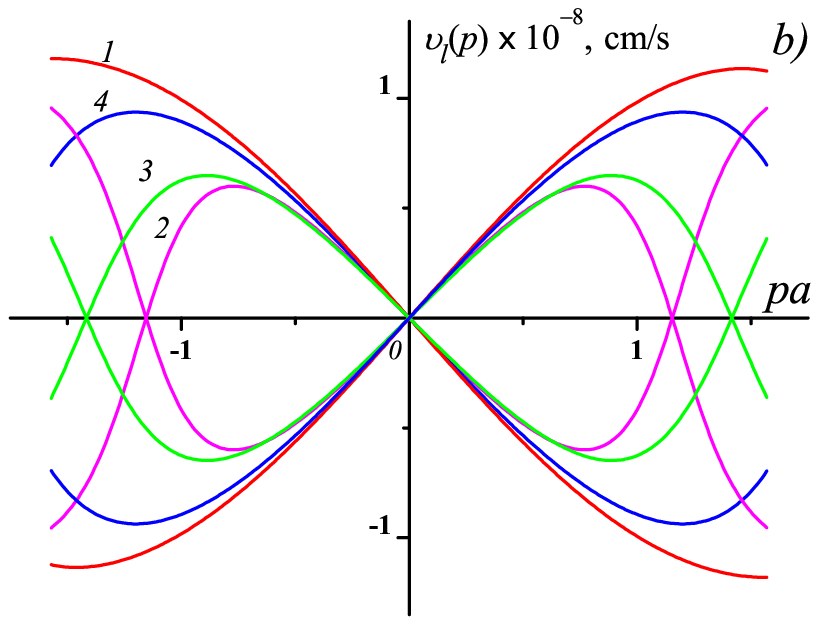}
\caption{Energy (a) and group velocity (b) vs quasi-momentum  for
(10,10) armchair nanotube. Numbers near curves are mode numbers
$l$.}\label{figdisp}
\end{figure}
Calculations of the energy and the group velocity for (10,10)
nanotube by Eqs. \eqref{RabA} and \eqref{grvel}  are presented  in
Fig. \ref{figdisp}. The curves in two figures can easily be
correlated: The larger slope of the dispersion curve the larger the
group velocity. Typical velocity of $\pi$-electrons excited to
energy of several electronvolts is about\cite{Reich_b04} $10^8$
cm/s. For such electrons, the synchronism condition requires the
electromagnetic wave slowing down as large as $300$ times. A proper
choice of the excited state in the vicinity of the group velocity
local maximums allows essential weakening this restriction.

The region in the vicinity of the  group velocity extremum is also
attractive because of the weak velocity dependence on the
quasi-momentum. As a result, in this region irradiation of photon
gets the electron only slightly out the synchronism condition
keeping high the probability to emit next photon. Due to that, the
radiation effectiveness grows in the vicinity of the group velocity
extremum. An additional advantage of  the local maximum in the group
velocity is the smaller negative influence of the beam energy spread
on the generation effectiveness. Indeed, in the vicinity of the
group velocity extremum the Taylor expansion of the energy does not
contain linear quasi-momentum terms. As a result, a larger number of
particles in a spreaded beam appears to be synchronized with
electromagnetic wave. This effect is characteristic for
quasi-particles and fully absent for free electrons.

The effect of radiation instability in nanotube  can be controlled
by  the variation of the electron effective mass.  The smaller the
mass the more responsive the electron is to perturbation, and the
more likely  an electron beam bunching. This means a faster
development of the instability. The reciprocal electron effective
mass is given by the quantity $\partial ^{2}\varepsilon
_{n}/\partial p_{n}^{2}$; therefore,  the increase of the
instability increment as the effective mass grows smaller follows
immediately from the dispersion equations (\ref{clasdisp}) and
(\ref{distribution}), which involve the reciprocal mass.

One more mechanism, which does not require large wave slowing down,
is exploiting electron interband transitions. In this case, as
follows from (\ref{finaldisp}), the resonance condition is
$\omega-v_{s}k=\Omega_{ns}$ (we suppose that transition frequency
exceeds the term related to the recoil) and the radiation frequency
can vary from infrared to ultraviolet. For interband transitions,
single--particle spontaneous emission of electron (positron) beams
emerging from outside into nanotube  was considered by Artru
\textit{et al.} \cite{Artru_05_PR}.

To weaken the requirement imposed  on the electromagnetic wave to be
slowed down to the electron velocity one can utilize the photon
diffraction on a periodic lattice of carbon atoms in a nanotube.
Resonance interaction takes place for harmonics corresponding to the
reciprocal vector $\tau$ satisfying the condition
$\omega-v_{n}(k+\tau)=0$. Then, taking into account the condition
$v_{n}/c \ll 1$, one can obtain:
\begin{equation}
\omega_\tau =\frac{\displaystyle \tau v_{n}}{\displaystyle
1-n_\mathrm{ref} v_{n}/c}. \label{spectr}
\end{equation}
Here $n_\mathrm{ref}=k c/\omega$ is the effective refractive index
of corresponding mode. The spatial period of a nanotube  varies in
wide range. For zigzag and armchair nanotubes it equals to $2.49$
\AA,~ while for chiral nanotubes the translation period achieves
$10$ nm and more depending on the nanotube indices. As a result, the
generated wavelength  varies from ultraviolet (for armchair and
zigzag CNTs) to infrared range  for nanotubes with translation
period $\sim  2 \sqrt{3} \pi R_{cn}$.

\section{Conclusion}
\label{conclusion}

In the present paper, aiming at the development of the physical
basis of a new class of nano-sized light sources, we have
investigated theoretically a recently proposed mechanism of the
generation of stimulated electromagnetic radiation by electron beam
in carbon nanotubes. The basic idea exploits an analogy between CNTs
and macroscopic electron devices and utilizes the effect of wave
slowing down in waveguides. Three basic properties of carbon
nanotubes, the strong slowing down of surface electromagnetic waves,
the ballisticity of the electron motion over typical CNT length, and
the extremely high electron current density reachable in CNTs, allow
proposing them as candidates for the development of nano-scale
Chernekov-type emitters for a wide frequency range from terahertz to
optical. The threshold conditions evaluated from the theoretical
model demonstrate that the development of CNT-based nanoFEL is
already feasible at realistic present-day parameters of nanotubes.

\acknowledgments

The research was partially supported by the INTAS project
05-1000008-7801, the EU FP7 TerACan project FP7-230778, the IB  BMBF
(Germany) project BLR 08/001,  and the Belarus Republican Foundation
for Fundamental Research project F08R-009.

\end{document}